\documentclass[conference]{IEEEtran}
\IEEEoverridecommandlockouts
\usepackage{cite}
\usepackage{amsmath,amssymb,amsfonts}

\usepackage{graphicx}
\usepackage{subcaption}
\usepackage{textcomp}
\usepackage{enumerate}
\usepackage{xcolor}
\usepackage{algorithm}
\usepackage{algpseudocode}
\usepackage{algpseudocodex}

\algrenewcommand\algorithmicrequire{\textbf{Input: }}
\algrenewcommand\algorithmicoutput{\textbf{Output: }}

\usepackage{soul}
\usepackage{tikz}

\def\BibTeX{{\rm B\kern-.05em{\sc i\kern-.025em b}\kern-.08em
    T\kern-.1667em\lower.7ex\hbox{E}\kern-.125emX}}

\newcommand{\Hermitian}{\mathrm{H}}

\newcommand{\bm}{\mathbf}
\newcommand{\be}{\begin{equation}}
\newcommand{\ee}{\end{equation}}
\newcommand{\bse}{\begin{subequations}}
\newcommand{\ese}{\end{subequations}}
\newcommand{\bea}{\begin{eqnarray}}
\newcommand{\eea}{\end{eqnarray}}
\newcommand{\x}{{\bm x}}

\newcommand{\bh}{{\bf h}}

\newcommand{\eye}{{\bm I}}

\begin{document}

\title{Windowed Dictionary Design for Delay-Aware OMP Channel Estimation under Fractional Doppler\\

\thanks{This publication has emanated from research conducted with the financial support of Science Foundation Ireland under Grant numbers SFI/19/FFP/7005(T) and SFI/21/US/3757, and National Science Foundataion under Grants 
ECCS-2153875 and CNS-2229562.   For the purpose of Open Access, the authors have applied a CC BY public copyright licence to any Author Accepted Manuscript version arising from this submission.
 }}
\author{\normalsize Hanning Wang$^*$, Xiang Huang$^\dagger$, Rong-Rong Chen$^\dagger$ and Arman Farhang$^*$  
\\$^*$Department of Electronic \& Electrical Engineering, Trinity College Dublin, Ireland, \\
$^\dagger$Department of Electrical and Computer Engineering, University of Utah, USA. \\
Email: \{wangh15, arman.farhang\}@tcd.ie, \{eric.xiang.huang, rchen\}@ utah.edu }

\maketitle

\begin{abstract}
Delay-Doppler (DD) signal processing has emerged as a powerful tool for analyzing multipath and time-varying channel effects. Due to the inherent sparsity of the wireless channel in the DD domain, compressed sensing (CS) based techniques, such as orthogonal matching pursuit (OMP), are commonly used for channel estimation. However, many of these methods assume integer Doppler shifts, which can lead to performance degradation in the presence of fractional Doppler. In this paper, we propose a windowed dictionary design technique while we develop a delay-aware orthogonal matching pursuit (DA-OMP) algorithm that mitigates the impact of fractional Doppler shifts on  DD domain channel estimation. First, we apply receiver windowing to reduce the correlation between the columns of our proposed dictionary matrix. Second, we introduce a delay-aware interference block to quantify the interference caused by fractional Doppler. This approach removes the need for a pre-determined stopping criterion, which is typically based on the number of propagation paths, in conventional OMP algorithm. Our simulation results confirm the effective performance of our proposed DA-OMP algorithm using the proposed windowed dictionary in terms of normalized mean square error (NMSE) of the channel estimate. In particular, our proposed DA-OMP algorithm demonstrates substantial gains compared to standard OMP algorithm in terms of channel estimation NMSE with and without windowed dictionary.
\end{abstract}


\section{Introduction}
Delay-Doppler (DD) signal processing has gained momentum in research on the design of future communication networks due to the reasons that are well articulated in \cite{OTFS,wei2021orthogonal,2022otfs,OTFSpredictability}.
DD domain multiplexing techniques, such as orthogonal time frequency space (OTFS) modulation \cite{OTFS}, leverage high resolution in both delay and Doppler domains in handling high-Doppler multipath channels. 
While time-domain channel impulse response rapidly varies in high-mobility scenarios, certain channel parameters, such as delay and Doppler shifts, remain relatively stable at least for a short period of time \cite{wei2021orthogonal}.    
Hence, the channel exhibits linear time-invariant (LTI) characteristics in the DD domain. Moreover, time-varying channels are known to have a sparse representation in the DD domain which potentially simplifies channel estimation \cite{OTFS}. However, accurate estimation of the channel gains, delay and Doppler shifts remains a significant challenge.

   Literature on DD domain channel estimation can be categorized into two classes. The first class, \cite{embedpilotcs, li2024reduced, bempilot}, is focused on finding the DD domain channel impulse response without separating the paths. 
   This is while the second class is concentrated on estimating the path gains and their corresponding delay and Doppler shifts that collectively form the discrete-time baseband channel response \cite{CSOMP1,4domp,2domp,subspacedoaforce}. {As  shown in \cite{4domp,chockalingam2020sparse} and \cite{underwatersparse}, compressed sensing (CS) is proven to be a powerful tool for estimating the channel parameters. Therefore, the main focus of this paper is on the second class of channel estimation techniques by compressed sensing.}

    Motivated by the quasi-static nature of the channel and its sparse representation in the DD domain, the authors in \cite{underwatersparse} applied matching pursuit (MP) techniques to estimate the delay and Doppler shifts for underwater channels. In \cite{chockalingam2020sparse},  orthogonal matching pursuit (OMP) was deployed to estimate the uplink channel parameters in OTFS where significant improvement over impulse-based channel estimation was reported. A structured 3D-OMP method with embedded pilot was proposed in \cite{CSOMP1} that exploits joint sparsity in delay-Doppler-angular (DDA) domain. Similarly, \cite{4domp} exploited sparsity in DDA domain using a 4D-OMP scheme, though with a different pilot and dictionary design compared to \cite{CSOMP1}. This framework was extended in \cite{2domp} where a simultaneous OMP (SOMP) method was proposed without considering the angular domain. The CS frameworks in \cite{CSOMP1,4domp,2domp} and \cite{chockalingam2020sparse} show performance gains and training overhead reduction compared to conventional methods, such as least squares (LS) channel estimation. However, all these works consider integer Doppler shifts, which may not be a realistic assumption. The performance of OMP-based algorithms depends heavily on the DD domain sparsity. Fractional Doppler shifts cause leakage across the Doppler dimension due to sampling point mismatch which leads to reduced sparsity \cite{interferencecancellation}.

    In \cite{OMPhastooversample}, the authors proposed an OMP-based scheme to deal with fractional Doppler. The main drawback of this scheme is that it requires a significantly large dictionary size. In \cite{CSOMP1,BEM} and \cite{subspaceomp}, the number of OMP iterations is chosen to be equal to the number of propagation paths which is not suitable for capturing the fractional Doppler effect. Alternatively, \cite{2domp} and \cite{4domp} set noise variance as a residue threshold for OMP. However, this method is limited to the scenario with integer Doppler shifts.

To address the aforementioned issues, in this paper, we propose a new windowed dictionary design technique and a modified OMP-based channel estimation scheme that accounts for fractional Doppler shifts. Our main contributions are summarized as follows:
\begin{itemize}
    
    \item {We propose a windowed, delay-aware (DA) dictionary design for estimating delay and fractional Doppler shifts. It incorporates a raised-cosine window to minimize frequency leakage and enhance orthogonality, while the dictionary structure features a novel interference block to enable delay-awareness. The results show that the windowed dictionary design outperforms the non-windowed design, particularly at higher signal-to-noise ratios.}

    \item {We propose the DA-OMP algorithm, in which interference is measured using a delay-aware interference block, without requiring prior knowledge of propagation paths, noise variance, or pre-determined thresholds. The stopping criteria is adaptively set to ensure reliable convergence.

    Simulation results demonstrate that the proposed DA-OMP achieves a channel estimation accuracy that is several orders of magnitude better than the standard OMP algorithm. Moreover, for the same oversampling factor in the dictionary matrix, DA-OMP achieves a significantly lower normalized mean square error in channel estimation compared to the OMP.}
\end{itemize}

The rest of the paper is structured as follows:  In Section~\ref{sec:system_model} we present the channel model. Section~\ref{sec:methodology} introduces the proposed windowed dictionary design and the interference block in the OMP. The simulation results are discussed in Section~\ref{sec:results}, and our conclusion is provided in Section~\ref{sec:conclusions}.

\textit{Notations:} In this paper, matrices, vectors and scalars are denoted as uppercase boldface, lowercase boldface letters and normal letters, respectively. $\mathbf{A}^{\rm T}$, $\mathbf{A}^{\Hermitian}$ and $\mathbf{A}^{\dagger}$ represent the transpose, Hermitian transpose and pseudo-inverse of the matrix, respectively. $\mathbf{A}^{:, \mathcal{I}}$ denotes the selection of columns in the set $\mathcal{I}$ and $\O$ represents the empty set. $a[i]$ denotes the $i$-th element of the vector $\mathbf{a}$. $\mathbf{I}_L$ denotes a $L \times L$ identity matrix. $\delta(\cdot)$ is the Dirac delta function. 
The function $\operatorname{diag}\{\mathbf{a}\}$ forms a diagonal matrix with the entrices of vector $\mathbf{\alpha}$ on the main diagonal. $\operatorname{rect}(\cdot)$ and $\operatorname{rcos}(\cdot)$ represents rectangular and raised-cosine window functions, respectively. $((\cdot))_L$ represents $L$-mod calculation. $\lfloor \cdot \rfloor$ and $\lceil \cdot \rceil$ represents floor and ceiling calculation, respectively.

\section{System Model} \label{sec:system_model}
We consider the discrete-time baseband transmit signal $\mathbf{s} \in \mathbb{C}^{L \times 1}$ with $L$ samples that are spaced every $\Delta\tau$ seconds apart.

The total duration and bandwidth of the transmit signal are $L\Delta \tau$ and $L \Delta \nu$, respectively, with $\Delta \tau$ and $\Delta \nu=\frac{1}{L\Delta \tau}$ denoting the delay and Doppler resolution. The DD domain multi-path channel response can be formulated as 
    \begin{equation}
        h(\tau, \nu)=\sum_{p=0}^{P-1} h_p \delta\left(\tau-\tau_p\right) \delta\left(\nu-\nu_p\right), 
    \label{eq:multipathchannel_dirac}
    \end{equation}
where $P$ is the total number of propagation paths, and $h_p$, $\tau_p$ and $\nu_p$ are the complex channel gain, delay, and Doppler shift of the $p^{\rm th}$ path, respectively.  After sampling the channel at the Nyquist rate, the normalized delay and Doppler shifts are denoted as
\begin{equation}
 \ell_p=\frac{\tau_p}{\Delta \tau}, \quad \kappa_p=\frac{\nu_p}{\Delta \nu},
\label{eq:ddgrid}
\end{equation}
where $\ell_p \in [0, \ell_{\max} - 1] $, $\kappa_p \in[0, \kappa_{\max}-1] $,
and $\ell_{\max}={\tau_{\max}}/{\Delta \tau}$ and $\kappa_{\max}={\nu_{\max}}/{\Delta \nu}$ are determined by the maximum delay spread and Doppler spread of the channel, i.e., $\tau_{\max}$ and $\nu_{\max}$, respectively. Assuming a high delay resolution, rounding the fractional delays to the nearest sample points is common practice \cite{Thaj2021}. Hence, we assume the delays $\ell_p$ take integer values and Doppler shifts $\kappa_p$ take either integer or fractional values. We add a cyclic-prefix (CP) with length $L_{\mathrm{cp}} \geq \ell_{\max}$ at the beginning of each block of $L$ samples to avoid inter-block interference (IBI). After CP removal, the received signal is represented as
\begin{equation}
r[n]=\sum_{p=0}^{P-1} h_p e^{j 2 \pi \frac{\kappa_p\left(n-\ell_p\right)}{L}} s\left[((n-\ell_p))_{L}\right]+\eta[n]
\label{eq:received_signal_series}
\end{equation}
with the vectorized form 
\begin{equation}
    \mathbf{r} = \mathbf{H} \mathbf{s} + \boldsymbol{\eta}, 
    \label{eq:received_signal_pre_window}
\end{equation}
where $\mathbf{r}=[r[0],\ldots,r[L-1]]^{\rm T}$ is the received signal vector, $\mathbf{s}=[s[0],\ldots,s[L-1]]^{\rm T}$ is the transmit signal vector,  and $\boldsymbol{\eta}\in\mathbb{C}^{L \times 1}$ is the noise vector, with the elements following a circularly symmetric  complex Gaussian distribution, i.e., $\eta[n] \sim \mathcal{CN}(0,\sigma^2)$. The delay-Doppler channel matrix, $\mathbf{H} \in \mathbb{C}^{L \times L}$, is constructed as in \cite{channelmodel1}
    \begin{equation}
\mathbf{H}=\sum_{p=0}^{P-1} h_p \mathbf{\Pi}^{ \ell_p}\mathbf{\Delta}^{\kappa_p} = \sum_{p = 0}^{P-1} h_p \boldsymbol{\Gamma}(\ell_p, \kappa_p),
\label{eq:channel_matrix_representation}
\end{equation}
where $\boldsymbol{\Gamma}(\ell_p, \kappa_p) = \mathbf{\Pi}^{ \ell_p}\mathbf{\Delta}^{\kappa_p}$ is the delay-Doppler shift component of the channel matrix for path $p$, $\boldsymbol{\Pi}^{\ell_p}$ denotes a $\ell_p$-step forward  cyclic shifted permutation matrix, and Doppler shift matrix $\boldsymbol \Delta ^ {\kappa_p}$ is structured as 
\begin{equation}
\boldsymbol{\Delta}^{\kappa_p}=\operatorname{diag}\left\{[1, e^{j \frac{2 \pi}{L}\kappa_p}, \ldots, e^{j \frac{2 \pi (L-1)}{L}{\kappa_p}}]^{\rm T}\right\}.
\label{eq:deltamat}
\end{equation}

\section{Proposed Dictionary Design}
\label{sec:methodology}
In this section, we first present our proposed pilot structure. Then, we define a DD domain grid that is oversampled along the Doppler dimension. Based on this grid and the proposed pilot structure, we develop a windowed dictionary design  to reduce the leakage that is caused by the fractional Doppler effect.
Therefore, our dictionary matrix is designed based on three key aspects: the pilot structure, on-grid channel representation, and windowing. 

\subsection{Proposed Pilot Structure}
\begin{figure}[t]
    \centering
    \includegraphics[width=\columnwidth]{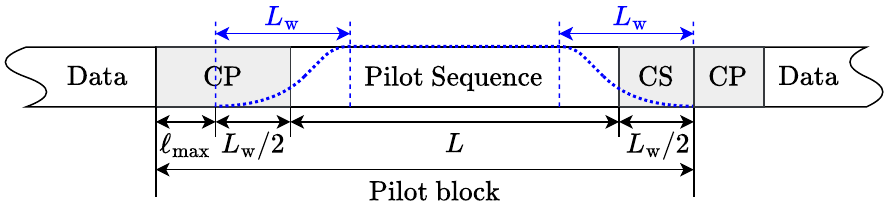}
    \caption{One frame of our proposed pilot structure. The blue curve represents the receiver time window with a roll-off length of $L_\mathrm{w}$.}
    \label{fig:pilot_structure}
\end{figure}

Let $\mathbf{x} \in \mathbb{C} ^ {L \times 1} $ denote the pilot sequence. 
As mentioned earlier, to avoid IBI, a CP is appended at the beginning of the pilot sequence. To accommodate time domain windowing in the dictionary design, our proposed pilot structure includes extra CP samples and a cyclic-suffix (CS) of length $L_\mathrm{cs}$ that is appended at the end of the sequence, see Fig.~\ref{fig:pilot_structure}. The pilot signal with cyclic extensions can be denoted as
\begin{equation}
    \overline{\mathbf{x}} =  \overline{\mathbf{T}} \mathbf{x}, 
    \label{eq:cyclic_extended_pilot}
\end{equation}
where $\overline{\mathbf{x}} \in \mathbb{C}^{L_\mathrm{tot} \times 1}$, $ \overline{\mathbf{T}} =\left[\mathbf{G}_{\mathrm{cp}}^{\rm T}, \mathbf{I}_L^{\rm T}, \mathbf{G}^{\rm T}_{\mathrm{cs}}\right]^{\rm T}$ is the cyclic extension matrix of size $L_\mathrm{tot} \times L$ and $L_\mathrm{tot} = L + L_\mathrm{cp} + L_\mathrm{cs}$. The matrices $\mathbf{G}_{\mathrm{cp}}$ and $\mathbf{G}_{\mathrm{cs}}$ include the last $L_\mathrm{cp}$ and the first $L_\mathrm{cs}$ rows of the identity matrix $\mathbf{I}_L$, respectively. 

\subsection{Proposed Dictionary Design}
Inspired by the works in \cite{zipper, cfowindowing, otfsglobalwindowing}, in this subsection, we propose to apply receiver windowing to limit the leakage originating from fractional Doppler shifts and oversampling.

Using \eqref{eq:received_signal_pre_window}, the received pilot sequence, after discarding its first $\ell_{\rm max}$ samples and windowing, can be represented as

\begin{equation}
    \widetilde{\mathbf{y}} = \mathbf{W}\mathbf{H}\widetilde{\mathbf{x}} + \widetilde{\boldsymbol{\eta}}
    \label{eq:dd_input_output}
\end{equation}
where $\widetilde{\x} = {\bf T} {\x}$,  ${\mathbf{T}} =\left[\mathbf{G}_{\mathrm{w}}^{\rm T}, \mathbf{I}_L^{\rm T}, \mathbf{G}_{\mathrm{cs}}^{\rm T}\right]^{\rm T}$  and $\mathbf{G}_{\mathrm{w}}$ is comprised of the last $L_{\rm cp}-\ell_{\rm max}$ rows of the identity matrix $\eye_L$.
$\mathbf{H} \in \mathbb{C}^{L' \times L'}$, $\mathbf{W} = \operatorname{diag}\{ \mathbf{w} \}$, $\mathbf{w}=[w[0],\ldots,w[L'-1]]^{\rm T}$, $L'=L_{\rm tot}-\ell_{\rm max}$ and $\widetilde{\boldsymbol{\eta}} = \mathbf{W} \boldsymbol{\eta}'$ where $\boldsymbol{\eta}' \sim \mathcal{CN}(0,\sigma^2\eye_{L'})$. In this paper, we consider raised-cosine window with the samples 
\begin{equation}
     w[n] = \operatorname{rect}\left( \frac{n-L/2}{L}\right) \ast c[n],
     \label{eq:raised_cosine_window}
\end{equation}
where
\begin{equation}
    c[n] = \frac{\pi}{2L_\mathrm{w}} \sin \left(\frac{\pi n}{L_\mathrm{w}}\right)\operatorname{rect}\left( \frac{n - L_\mathrm{w}}{L_\mathrm{w}}\right),
    \label{eq:raised_cosine_window_cn}
\end{equation}
for $n=0,\ldots,L'-1$ and $L_\mathrm{w}$ samples in the roll-off period as demonstrated in Fig.~\ref{fig:pilot_structure}. When $L_\mathrm{w} = 0$, $\mathbf{w}$ reduces to the rectangular window, and $\mathbf{W}$ becomes the identity matrix.

Next, we define an oversampled DD grid along the Doppler dimension for off-grid approximation. As observed in \eqref{eq:multipathchannel_dirac} and \eqref{eq:channel_matrix_representation}, the DD domain channel is a combination of the multipath components. Thanks to the sparse representation of the channel in DD domain \cite{wei2021orthogonal}, block length can be chosen to be much larger than the number of paths, i.e. $ L \gg P$. Thus, we can exploit the DD domain sparsity to efficiently estimate the delay and Doppler shifts of the channel. To this end, we form a DD grid following the approach in \cite{4domp}, but with the consideration of oversampling along the Doppler dimension

\begin{equation}
\mathcal{G}(l, k) = \left\{ l, \ k \,\middle|\, 
\begin{array}{l}
0 \leq l \leq G_\tau - 1, \\
0 \leq k \leq G_\nu - 1 
\end{array}
\right\}
\label{eq:dic_dd_grid}
\end{equation}
where $l$ and $k$ are integer delay and Doppler indices on the grid, 
\begin{equation}
G_\tau = \lceil \frac{\tau_{\max}} { \Delta \tau} \rceil,  \quad G_\nu = \lceil u_\nu \frac{\nu_{\max}}{ \Delta \nu} \rceil,
\label{eq:g_tau_and_g_nu}
\end{equation}
where $u_\nu$ is the oversampling factor along the Doppler dimension that defines the Doppler resolution of the grid as $\Delta\nu/u_\nu$. 
In presence of fractional Doppler shifts, oversampling is a necessity for obtaining accurate channel estimates.

Next, we formulate the channel estimation problem as a sparse recovery problem, i.e., 
\bea
    \widetilde{\mathbf{y}} 
=\mathbf{\Phi}_\mathrm{S}\widehat{\mathbf{h}} + \widetilde{\boldsymbol{\eta}},
 \label{eq:dic_input_output_for_omp}
\eea
where 
$\widehat{\bh} \in \mathbb{C}^{G_\tau G_\nu \times 1}$ is a sparse vector and 
$\boldsymbol{\Phi}_\mathrm{S} \in \mathbb{C}^{L' \times G_\tau G_\nu}$ is our proposed windowed dictionary matrix.  The construction of $\boldsymbol{\Phi}_\mathrm{S}$ is motivated by the structure of the channel matrix in (\ref{eq:channel_matrix_representation}) and the received signal model in (\ref{eq:dd_input_output}). Specifically,   
we map each pair of the delay-Doppler indices $(l, k)$, where $l\in\{0,\ldots, G_\tau -1\}$ and $k \in \{0,\ldots,G_\nu -1\}$, 
 to a column of $\boldsymbol{\Phi}_\mathrm{S}$, denoted by  $\boldsymbol{\phi}_{d}$, 
where $d=l G_{\nu}+k$. We define the $d$-th column of the dictionary matrix as
  $\boldsymbol{\phi}_{d} =  \mathbf{W} \widetilde{\boldsymbol{\Gamma}}(l, k) \widetilde{\mathbf{x}}$, where
$\widetilde{\boldsymbol{\Gamma}}(l, k) = \mathbf{\Pi}^{ l}\widetilde{\mathbf{\Delta}}^{k}$ represents a component channel matrix resulting from the $l$-th delay and $k$-th Doppler shift defined on the oversampled grid, and 
\begin{equation}
\widetilde{\boldsymbol{\Delta}}^{k}\!=\!\operatorname{diag}\!\left\{\!\left[e^{j\frac{-2 \pi (L_\mathrm{w}/2)}{ Lu_\nu}k}, \ldots, 1, \ldots, e^{j\frac{2 \pi (L'-1)}{L u_\nu }k} \right]^{\rm T}\!\right\}.
\label{eq:deltamat_oversampled_on_grid}
\end{equation}
Using the DA-OMP algorithm presented in the following section, we will obtain a sparse solution $\widehat{\mathbf{h}}$ to (\ref{eq:dic_input_output_for_omp})  which has $Q$ non-zero components with values $\widehat{h}_q$ on its entries $d_q = l_q G_\nu + k_q$ for $q=0,\ldots,Q-1$.
As a result,  the actual channel matrix $ \mathbf{H}$ can be approximated by a channel matrix defined on the grid, $\widetilde{\mathbf{H}}$, where 
\begin{equation}
   \widetilde{\mathbf{H}} = \sum_{q = 0}^{Q-1} \widehat{h}_q \widetilde{\boldsymbol{\Gamma}}(l_q, k_q).
    \label{eq:on_grid_channel}
\end{equation}

Based on the above explanations, our proposed dictionary matrix has the following block structure
\begin{equation}
\boldsymbol{\Phi}_\mathrm{S}=\left[\boldsymbol{\Phi}_{0}, \ldots, \boldsymbol{\Phi}_{l}, \ldots, \boldsymbol{\Phi}_{G_\tau-1}\right],
\label{eq:window_block_dic}
\end{equation}
with the columns of each block $\boldsymbol{\Phi}_{l}=[\boldsymbol{\phi}_{l G_\nu},\ldots,\boldsymbol{\phi}_{(l+1)G_\nu-1}]$ corresponding to the same delay value $l$, and the Doppler shifts $k \in\{0,\ldots,G_\nu - 1\}$.

\section{Proposed Delay-aware OMP-based \\ Channel Estimation}\label{sec:DA-OMP}

In this section, we propose a delay-aware OMP (DA-OMP) algorithm used to solve the sparse recovery problem in (\ref{eq:dic_input_output_for_omp}). The DA-OMP is specifically designed for the fractional Doppler scenario where the conventional OMP algorithm suffers from both the leakage problem and the inaccuracy in the terminating criterion.  

The OMP algorithm provides an efficient way to solve  sparse recovery problems like \eqref{eq:dic_input_output_for_omp}. In the integer Doppler case, the actual channel aligns perfectly with the grid. The number of peaks in the correlation value, $|\boldsymbol{\Phi}_\mathrm{S}^{\Hermitian}\widetilde{\mathbf{y}}|$, and the non-zero elements in $\widehat{\bh}$ directly correspond to the number of propagation paths $P$. The number of iterations for OMP then can be set based on the knowledge of $P$ \cite{CSOMP1,subspacedoaforce,BEM}.
When Doppler shifts are integer, authors in \cite{4domp} and \cite{2domp} use the noise variance as a threshold to prevent OMP from searching under the noise floor. 
 However, with fractional Doppler shifts, channel $\mathbf{H}$ is off-grid, $\mathbf{p}_\mathrm{S}$ typically exhibits more than $P$ peaks due to leakage. 
 
  There are two key features in the DA-OMP. First, we adopt receiver filtering to reduce the leakage in the fractional Doppler case. Second, we  introduce the concept of an interference block, appended at the end of the dictionary, The DA-OMP measures  the interference using this interference block, and adaptively sets the stopping criteria to ensure convergence.

The proposed DA-OMP is summarized  in  Algorithm~\ref{alg:omp2}. 
 With our windowed  dictionary given in \eqref{eq:window_block_dic}, 
the $d$-th entry of $\mathbf{p}_\mathrm{S}$, where $d=l G_{\nu}+ k$, is 
\begin{equation}
\boldsymbol{\phi}_{{l},k}^\Hermitian \widetilde{\mathbf{y}}=\mathbf{x}^\Hermitian \mathbf{T}^\Hermitian \widetilde{\boldsymbol{\Gamma}}(l, k)^\Hermitian \widetilde{\mathbf{W}} \mathbf{H} \mathbf{T} \mathbf{x},
\end{equation}

where $\widetilde{\mathbf{W}} = \mathbf{W}^\Hermitian \mathbf{W}$.  For simplicity, we consider $\widetilde{\mathbf{W}}$ as one single raised-cosine window. The windowing can effectively reduce the leakage caused by the mismatch of on-grid $\widetilde{\boldsymbol{\Gamma}}(l, k)$ and off-grid ${\bf {H}}$ \cite{cfowindowing}.

\addtolength{\topmargin}{0.01in}
\begin{figure}[t]
    \centering
    \includegraphics[width=0.6\columnwidth]{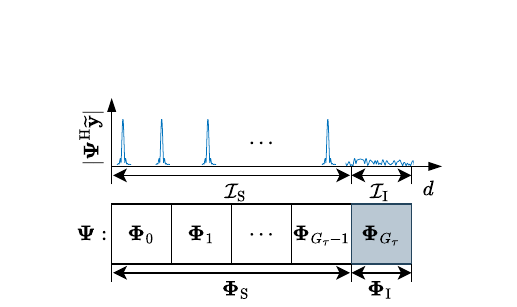}
    \caption{Delay-aware dictionary design for DA-OMP. }
    \label{fig:interference_block}
    \vspace{-0.5cm}
\end{figure}

After leakage reduction, fractional Doppler values can be approximated by a few adjacent points near the on-grid peak\cite{interferencecancellation}. As a result, additional iterations and alternative stopping criterion are necessary to capture the effects of fractional Doppler shifts. The motivation of the proposed DA-OMP is to ensure that the algorithm will terminate as soon as the search falls into the interference region.

Next, we propose an interference-based stopping criterion which will terminate the DA-OMP according to the measured interference at each iteration. This is achieved by appending an interference block $\boldsymbol{\Phi}_\mathrm{I}$ at the end of $\boldsymbol{\Phi}_\mathrm{S}$. The block $\boldsymbol{\Phi}_\mathrm{I}$ follows the same structure of a single block in \eqref{eq:window_block_dic} but with a larger delay of  ${l} = G_\tau$, where $G_\tau$ is the maximum index value along delay dimension on the grid as defined in \eqref{eq:g_tau_and_g_nu}. The overall dictionary now becomes
\begin{equation}
    \boldsymbol{\Psi} = \left[ \boldsymbol{\Phi}_\mathrm{S} \  \boldsymbol{\Phi}_\mathrm{I}\right],
    \label{eq:delay_aware_dic}
\end{equation}
and the columns of $\boldsymbol{\Phi}_\mathrm{I}$ are indexed by $\mathcal{I}_\mathrm{I}= \{G_\tau G_\nu, G_\tau  G_\nu  + 1, \cdots,   (G_\tau +1) G_\nu -1 \},$ corresponding to the fixed delay of 
$l=G_\tau$. 
On this basis, our proposed dictionary in \eqref{eq:delay_aware_dic} is delay-aware.  As shown in Fig.~\ref{fig:interference_block}, the peaks in $|\boldsymbol{\Psi}^\mathrm{H}\widetilde{\mathbf{y}}|$ always fall within $\mathcal{I}_\mathrm{S} = \{ 0, 1, \ldots, G_\tau G_\nu - 1\}$, which includes all column indices in $\boldsymbol{\Phi}_\mathrm{S}$. Conversely, $\mathcal{I}_\mathrm{I} $ contains column indices in $\boldsymbol{\Phi}_\mathrm{I}$, representing only interference in $\mathbf{p}$ as the actual channel $\mathbf{H}$ does not contain any delay tap at $\ell_p \geq \ell_{\max} = G_\tau$. We note that choosing $l$ to be any other values larger than $G_\tau$ does not impact the estimation performance. Hence,  we simply set $l = G_\tau$ for $\boldsymbol{\Phi}_\mathrm{I}$. By doing this, we can explicitly measure the maximum interference level in each iteration
\begin{equation}
\gamma_{i+1} = \underset{ j \in \mathcal{I}_\mathrm{I} }{\max } \left|\left(\boldsymbol{\Psi}^{:,j}\right)^\Hermitian \mathbf{r}_{i+1}\right|,
    \label{eq:max_interference}
\end{equation}
where $i$ denotes the $i$-th round of iteration, and $\mathbf{r}_{i+1}$ is the residue for the next iteration, with $\mathbf{r}_{0} = \widetilde{\mathbf{y}}$. If 
\begin{equation}
    \beta_i = \underset{ j \in \mathcal{I}_\mathrm{S} } {\max }\left|\left({\boldsymbol{\Psi}^{:, j}}\right)^\Hermitian \mathbf{r}_{i}\right| < \gamma_i,
    \label{eq:omp_stopping_criterion}
\end{equation}
indicates that, in $i$-th round, the maximum correlation value  $\beta_i$ is smaller than interference value. The DA-OMP should stop to prevent selection from the interference, to ensure lower estimation error and faster convergence.

\addtolength{\topmargin}{0.02in}
\begin{algorithm}[t]
\caption{Delay-aware OMP Channel Estimation}\label{alg:omp2}
\begin{algorithmic}[1]
\Require Dictionary $\boldsymbol{\Psi}$, measurement $\widetilde{\mathbf{y}}$, index sets $\mathcal{I}_\mathrm{S}$, $\mathcal{I}_\mathrm{I}$
\State  $\mathcal{I}_\mathrm{OMP} = \O$, $i = 0$,  $\mathbf{r}_0=\widetilde{\mathbf{y}}$, $\gamma_{0} = 0$, 

$\beta_0=\underset{ j \in \mathcal{I}_\mathrm{S} } {\max }\left|\left({\boldsymbol{\Psi}^{:, j}}\right)^{\rm H} \mathbf{r}_{0}\right|$

\While {$\beta_i = \underset{ j \in \mathcal{I}_\mathrm{S} } {\max }\left|\left({\boldsymbol{\Psi}^{:, j}}\right)^{\rm H} {\mathbf{r}_{i}}\right| > \gamma_{i}$} 
    \State ${d}_i=\underset{j \in \mathcal{I}_\mathrm{S} }{\arg \max } \ \left|\left({\boldsymbol{\Psi}^{:, j}}\right)^{\rm H} {\mathbf{r}_{i}}\right|$ 
    \State $\mathcal{I}_\mathrm{OMP}=\mathcal{I}_\mathrm{OMP} \cup {d}_i$
    \State $\widehat{\bh}_i=\left(\boldsymbol{\Psi}^{:, \mathcal{I}_\mathrm{OMP}}\right)^{\dagger} \widetilde{\mathbf{y}}$
    \State ${\mathbf{r}_{i+1}}=\widetilde{\mathbf{y}} -{\boldsymbol{\Psi}}^{:, \mathcal{I}_\mathrm{OMP}} \widehat{\bh}_i$ 

    \State ${\gamma_{i+1}} = \underset{ j \in \mathcal{I}_\mathrm{I} }{\max } \left|\left(\boldsymbol{\Psi}^{:,j}\right)^{\rm H} {\mathbf{r}_{i+1}}\right|$
    \State {$i = i + 1$}
\EndWhile

\State \Output $Q = i - 1$ and $\widehat{\bh}_{\mathrm{OMP}}=\widehat{\bh}_{i-1}$

\State Reconstruct estimated $\widetilde{\mathbf{H}}$ using \eqref{eq:on_grid_channel}.
\end{algorithmic}
\end{algorithm}

\section{Simulation Results} \label{sec:results}

In our simulations, we consider $\x$ as a pseudo-noise pilot sequence with length $L = 128$ that is chosen from binary phase-shift keying (BPSK) alphabet. The channel gains $h_p$ for $p=0,\ldots,P-1$ take random values between 0 and 1, and the number of paths $P$ is uniformly distributed between 5 and 8. The path delays, $\ell_p $, are uniformly distributed on the grid within the range $[0, G_\tau - 1]$, while the Doppler shifts $\kappa_p$ are uniformly generated in the range $[0, G_\nu - 1]$ with both integer and fractional parts, i.e., off-grid Doppler shifts. The normalized mean square error (NMSE) is calculated as 
\begin{equation}
    \mathrm{NMSE} = \mathbb{E}\left\{\frac{\left\|\widetilde{\mathbf{H}}-\mathbf{H}\right\|_2^2}{\left\|\mathbf{H}\right\|_2^2}\right\}.
    \label{eq:NMSE_Cal}
\end{equation}
where $\mathbb{E}\{\cdot\}$ represents the expected value. 

\begin{figure}[t]
    \begin{minipage}{0.49\columnwidth}
    \includegraphics[width=1\linewidth]{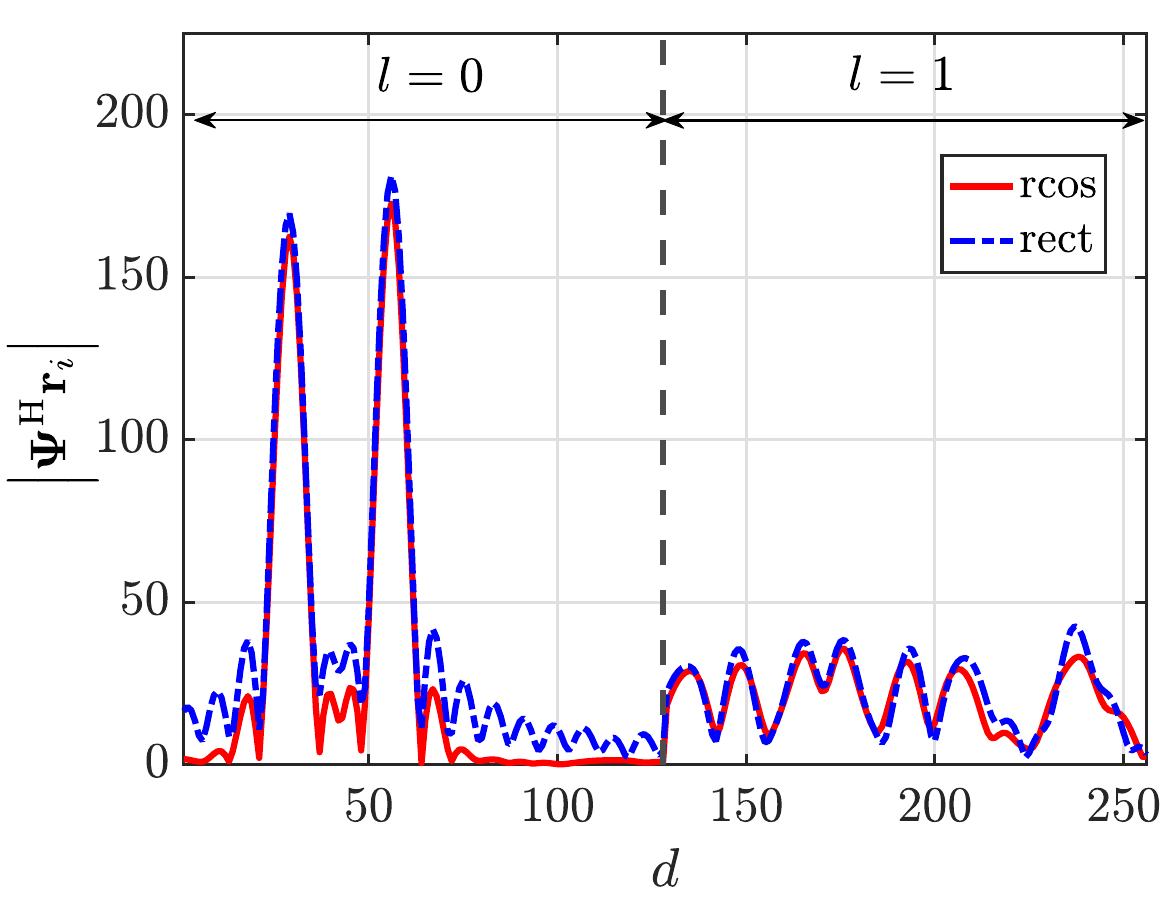}
    \subcaption{$i=0$}
    \label{fig:interference_block_i_0}
    \end{minipage}
    \begin{minipage}{0.49\columnwidth}
    \includegraphics[width=1\linewidth]{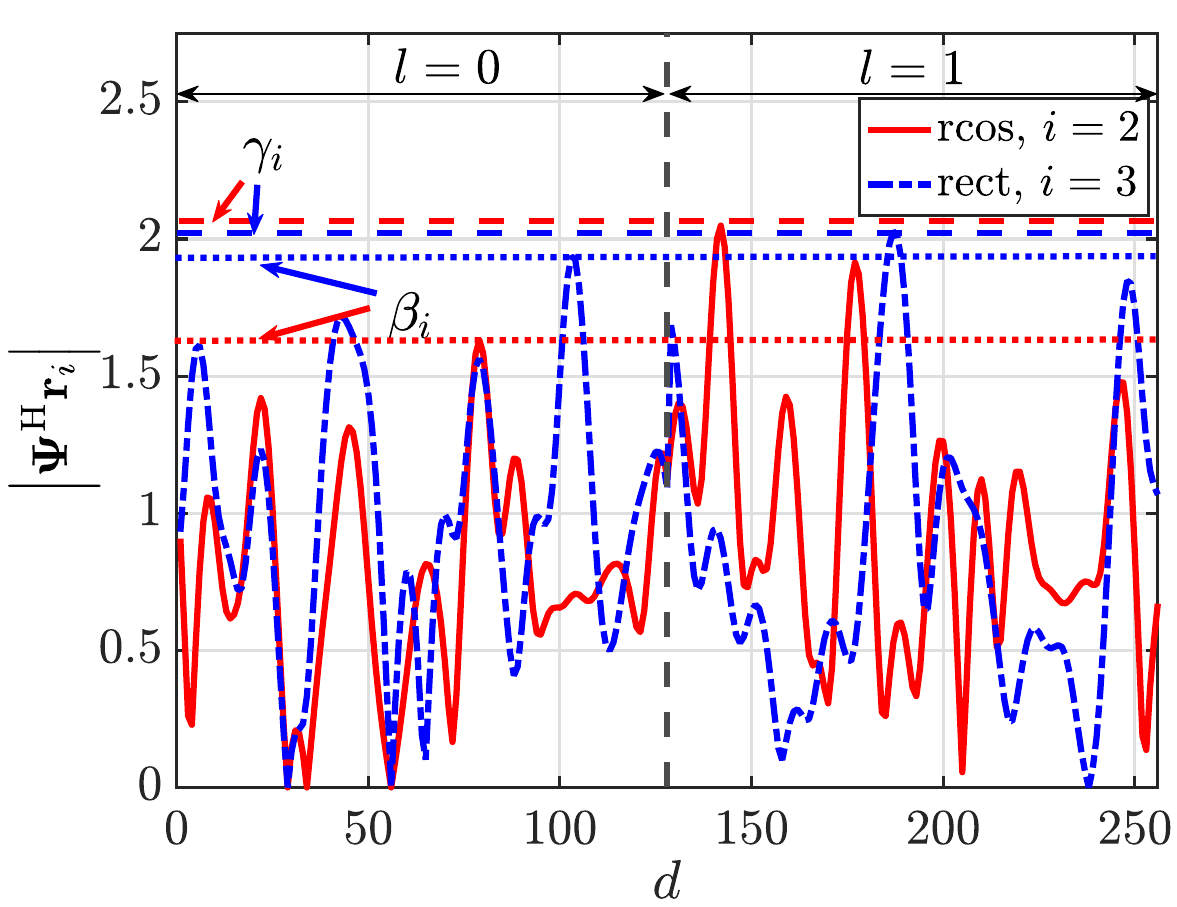}
    \subcaption{$i = 2$ for rcos, $i = 3$ for rect}
    \label{fig:interference_block_i_2_3}
    \end{minipage}
    
    \caption{The
    correlation values $\mathbf{p}_i$ as a function of the dictionary column index $d$
    at the $i$-th iteration of DA-OMP}
    \label{fig:interference_block_simulation}
    \vspace{-0.5cm}
\end{figure}

\begin{figure}[t!]
    \centering
    \includegraphics[width = 0.9\columnwidth]{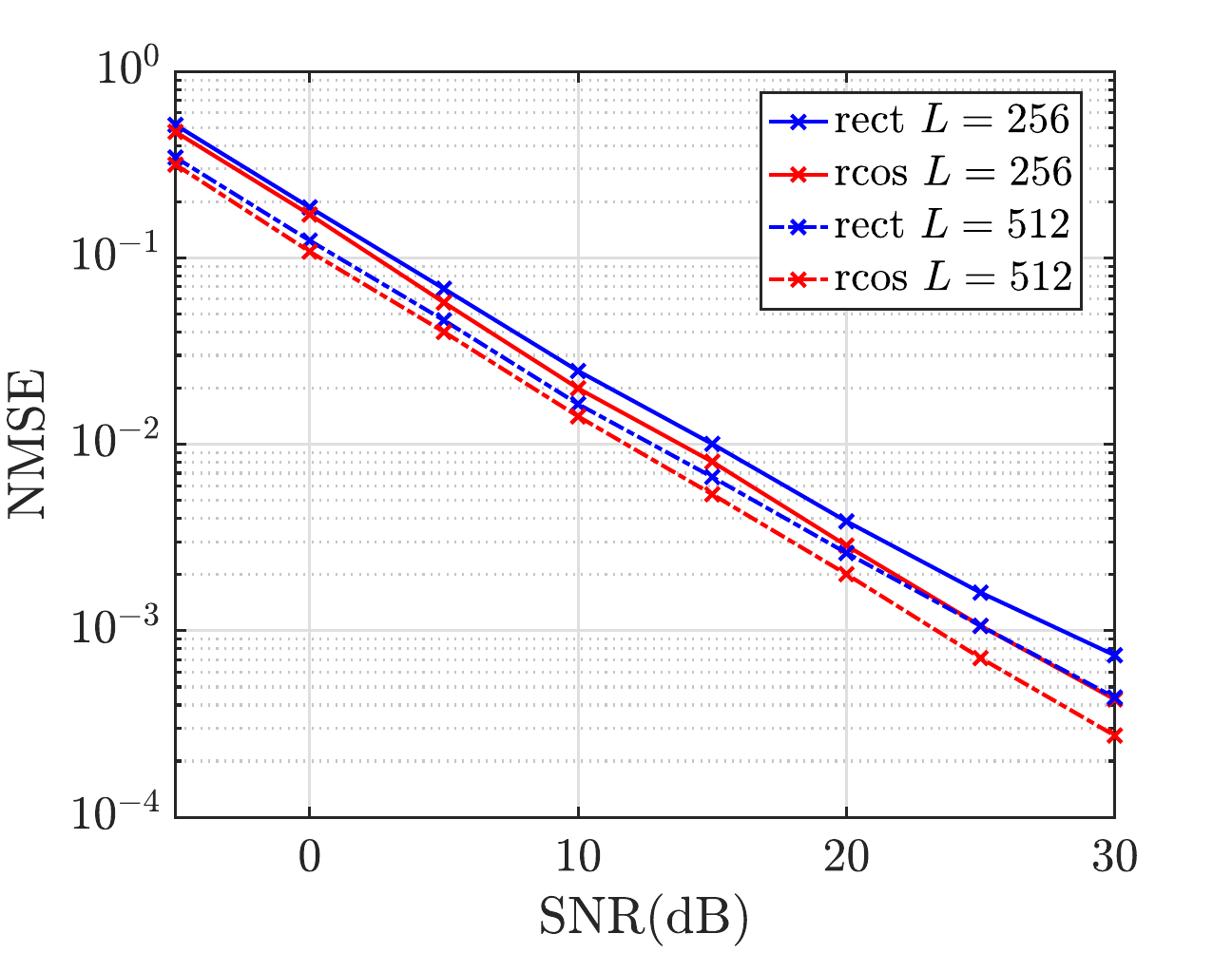}
    \caption{NMSE comparison for different values of $L$ using DA-OMP, with $L_\mathrm{w} = L/4$, $G_\tau = 4$, $G_\nu = 16$.}
    \label{fig:M_N_comp}
\end{figure}

In Fig.~\ref{fig:interference_block_simulation}, we plot the correlation vector $\left|\boldsymbol{\Psi}^\Hermitian{\mathbf{r}_i}\right|$ with and without receiver  windowing.
In Fig.~\ref{fig:NMSE_vs_taps}, we compare the NMSE performance of DA-OMP and OMP for different values of $G_\tau$. In Fig.~\ref{fig:Lw_comp}, we investigate the impact of roll-off length $L_\mathrm{w}$ on the NMSE performance. Finally, in Figs.~\ref{fig:kappa_comp} and \ref{fig:M_N_comp}, we examine the impact of the oversampling factor  $u_\nu$ and the pilot sequence length $L$ on the  NMSE performance, respectively. We use “rect” to represent the dictionary design without windowing, “rcos” for the dictionary design with raised-cosine windowing, “OMP” for the standard OMP, and “DA-OMP” for the delay-aware OMP. The key observations are summarized as follows.
\begin{itemize}
\item {In Fig.~\ref{fig:interference_block_i_0}, under $\mathrm{SNR} = 20~\mathrm{dB}$ with $G_\tau = 1$ and $G_\nu = 128$, the channel has $2$ paths with the same delay of $l=0$, but with different Doppler shifts. Two dominant peaks of the correlation vector at the first iteration, $|\boldsymbol{\Psi}^\Hermitian {\mathbf{r}_0}|$, appear only in the first half, corresponding to indices in $\mathcal{I}_\mathrm{S}=\{0, 1, \cdots, G_{\nu}-1=127\}$. After applying the raised-cosine window with roll-off length $L_\mathrm{w} = L/2$, the fractional leakage is significantly reduced. In the second half, with column indices in $\mathcal{I}_\mathrm{I}=\{G_\nu=128, \cdots,  (G_\tau+1)G_{\nu}- 1=255\}$, where the interference block $\boldsymbol{\Phi}_\mathrm{I}$ is generated with $l = 1$. The interference level increases due to the mismatched delay values. Fig.~\ref{fig:interference_block_i_2_3} indicates that DA-OMP terminates at $i = 2$ for the raised-cosine window and at $i = 3$ for the rectangular window, as $\beta_i$ falls below $\gamma_i$ in each case.}
\item In Fig.~\ref{fig:NMSE_vs_taps}, with $G_\nu = 16$, $G_\tau \!\in\!\{ 1, 4\}$, $u_\nu = 2$, $L_\mathrm{w} = L/2$, 
the results show that windowing improves the performance of both OMP and DA-OMP. The maximum gain is achieved by DA-OMP, reaching around 5~dB performance improvement at NMSE of $10^{-3}$ for dictionary with `rcos' compared to `rect'. Additionally, while the standard OMP fails to capture the effects of fractional Doppler shifts and hits a performance floor, DA-OMP consistently achieves lower NMSE. This demonstrates the robustness of DA-OMP in doubly-selective channels with fractional Doppler shifts. 

    \begin{figure}
    \centering
    \vspace{-0.4cm}
    \includegraphics[width = 0.9\columnwidth]{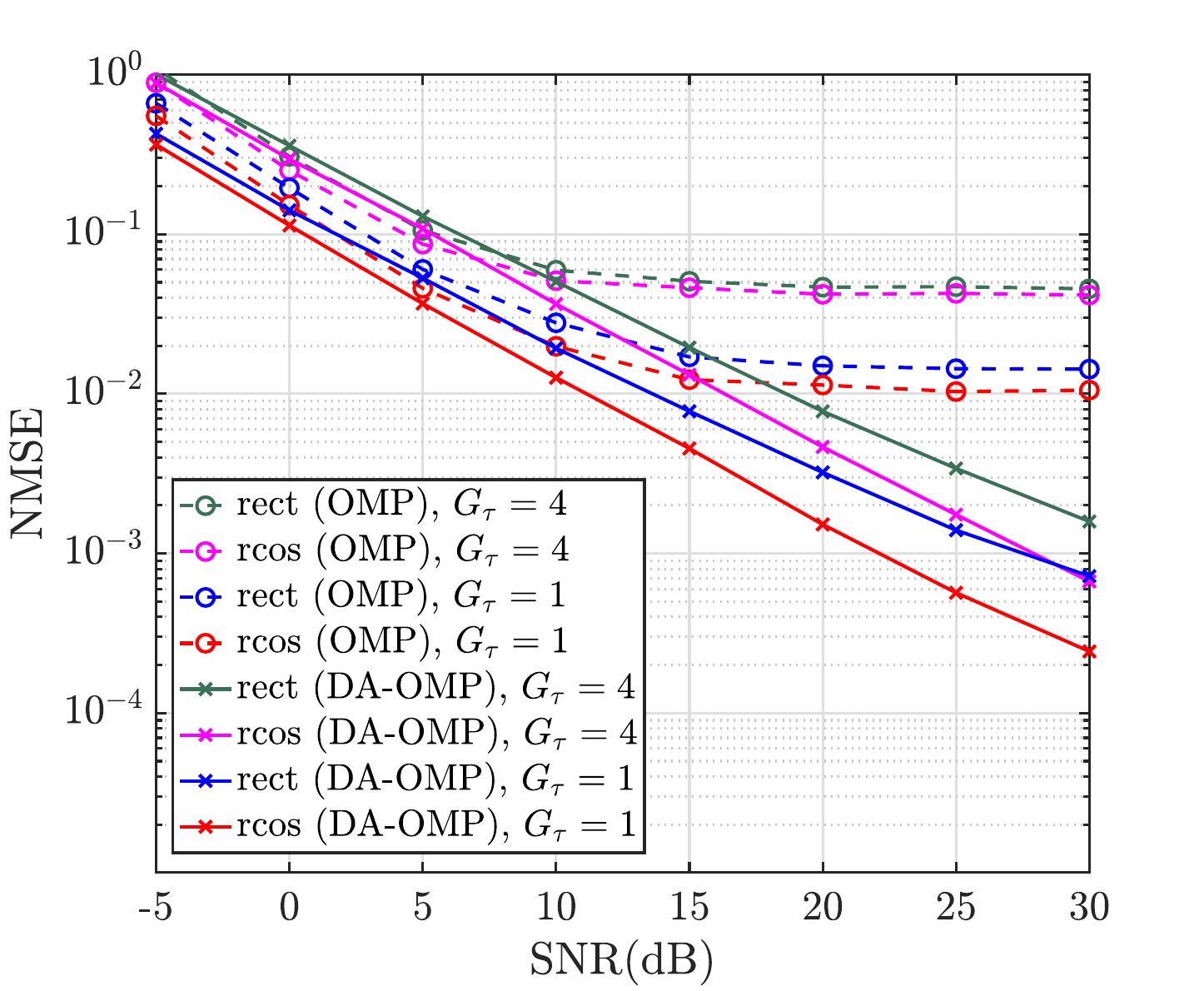}
    \vspace{-0.2cm}
    \caption{NMSE vs. SNR comparison of DA-OMP and OMP, with different values of $G_\tau$. }
    \label{fig:NMSE_vs_taps}
\end{figure}

    \begin{figure}
    \centering
    \includegraphics[width = 0.9\columnwidth]{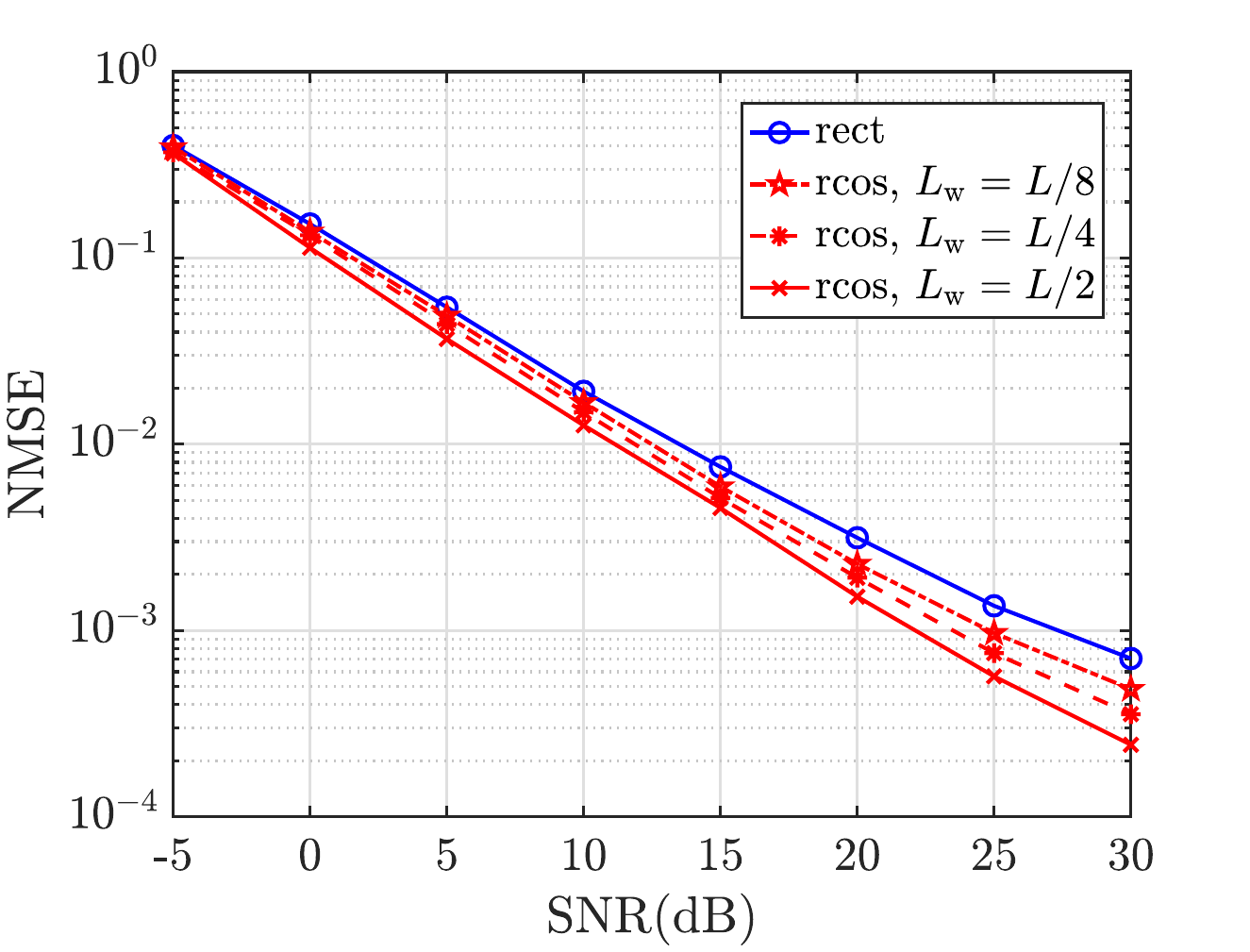}
    \caption{Impact of $L_\mathrm{w}$ on NMSE for DA-OMP for $G_\tau = 1$, $G_\nu = 16$ and $L=128$.}
    \label{fig:Lw_comp}
\end{figure}

    \begin{figure}
    \centering
    \includegraphics[width = 0.9\columnwidth]{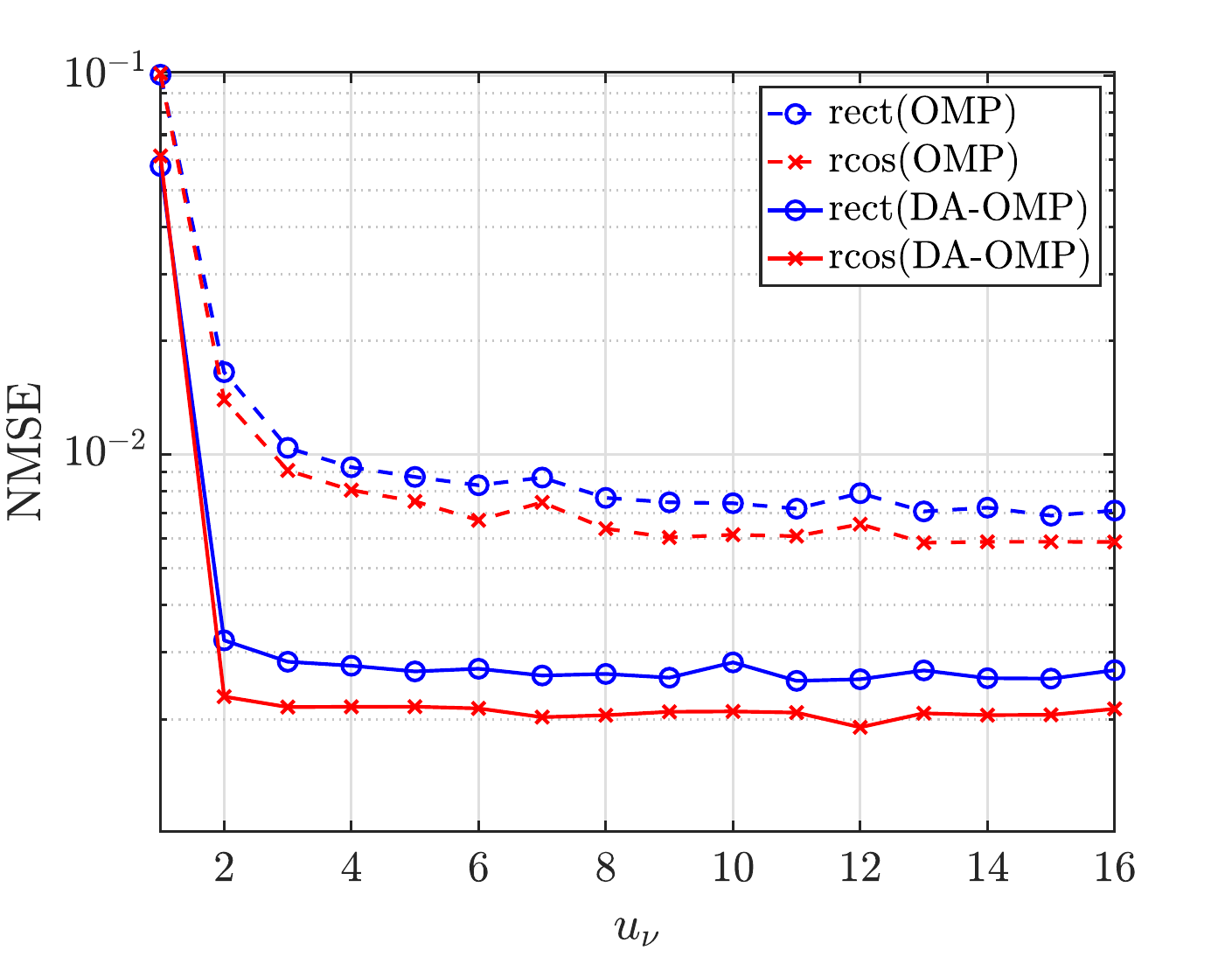}
    \caption{NMSE vs. $u_\nu$ for  $G_\tau = 1$, $G_\nu = 16$, $L=128$ and $\mathrm{SNR} = 20~\mathrm{dB}$. }  
    \label{fig:kappa_comp}
\end{figure}

\item In Fig.~\ref{fig:Lw_comp}, the results show that increasing $L_\mathrm{w}$ from $L/8$ to $L/2$ improves the NMSE performance for the windowed DA-OMP approach, especially at high SNRs. However, this comes at the cost of additional spectral resources. Hence, the trade-off between estimation accuracy and spectral efficiency should be carefully considered when choosing $L_\mathrm{w}$. 

\item In Fig.~\ref{fig:kappa_comp}, we fix $L = 128$, and plot   the oversampling factor 
 $u_\nu$ against NMSE. A larger $u_\nu$ increases the number of columns in $\boldsymbol{\Psi}$, which directly impacts computational complexity. The results show that with $\mathrm{SNR}= 20~\mathrm{dB}$, the DA-OMP consistently achieves a lower NMSE than those of the standard OMP with smaller $u_\nu$ values. Thus, DA-OMP offers an improved estimation accuracy with a lower computational complexity than OMP. In this figure, we also observe a performance floor as $u_\nu$ increases. The error floor is primarily influenced by the Doppler resolution, $\Delta \nu=\frac{1}{L\Delta \tau}$, which depends on the pilot length $L$. In other words, a higher Doppler resolution is not achieved by increasing the oversampling factor, $u_\nu$.

\item Fig.~\ref{fig:M_N_comp} further illustrates the impact of pilot length $L$ on NMSE, showing performance improvements with larger $L$ leading to a higher Doppler resolution. Notably, for $L=256$, windowing with $L_\mathrm{w} = L/4$ achieves comparable performance to no-windowing when $L=512$. This demonstrates that windowing not only improves estimation accuracy but also reduces the pilot overhead.

\end{itemize}

\section{Conclusion} \label{sec:conclusions}

In this work, we proposed a windowed dictionary design for delay-domain  channel estimation under fractional Doppler shifts. The proposed DA-OMP method introduces a novel interference block to the dictionary matrix, enabling an adaptive stopping criterion based on interference measurements. This eliminates the need for fixed threshold, which is typically required in other OMP approaches. Simulation results demonstrate that the proposed method achieves superior channel estimation accuracy compared to those of the standard OMP, both with and without windowing. Furthermore, we show that windowing achieves comparable performance with shorter pilot sequences, effectively reducing pilot overhead.

\bibliographystyle{IEEEtran}
\bibliography{biblio}

\end{document}